\documentclass[12pt]{article}
\usepackage{amsmath,amssymb,amsfonts}
\usepackage{url}
\usepackage{epsfig}
\usepackage[paper=letterpaper,margin=1.0in]{geometry}

\newcommand{\be}{\begin{equation}}
\newcommand{\ee}{\end{equation}}
\newcommand{\bea}{\begin{eqnarray}}
\newcommand{\eea}{\end{eqnarray}}
\newcommand{\ba}{\begin{array}}
\newcommand{\ea}{\end{array}}
\newcommand{\nn}{\nonumber}
\newcommand{\tr}{\mathrm{tr}\,}
\newcommand{\IC}{\mathbb{C}}
\newcommand{\IP}{\mathbb{P}}
\newcommand\cg{\mathfrak{g}}
\newcommand{\cM}{{\cal M}}
\newcommand{\cN}{{\cal N}}
\newcommand{\cF}{{\cal F}}
\newcommand{\eB}{{\bar{e}}}
\newcommand{\yB}{{\bar{y}}}
\def\eps{\epsilon^{\alpha \beta}}
\def\barH{\overline{H}}

\begin{document}

\thispagestyle{empty}

{\footnotesize
\rightline{WITS-CTP-129}
}

\renewcommand{\thefootnote}{\fnsymbol{footnote}}
\vskip1.5cm
\begin{center}
{\Large \textbf{Veronese Geometry and\\\vspace{0.25cm} the Electroweak Vacuum Moduli Space}}

\vskip1.0cm

\textbf{
Yang-Hui He${}^{a,b,}$\footnote{\texttt{hey@maths.ox.ac.uk}},
Vishnu Jejjala${}^{c,}$\footnote{\texttt{vishnu@neo.phys.wits.ac.za}},
Cyril Matti${}^{a,}$\footnote{\texttt{Cyril.Matti.1@city.ac.uk}},
Brent D.\ Nelson${}^{d,}$\footnote{\texttt{b.nelson@neu.edu}}}

\vskip0.5cm

${}^a$\textit{Department of Mathematics, City University, London,\\
Northampton Square, London EC1V 0HB, UK}
\vskip0.25cm

${}^b$\textit{School of Physics, NanKai University, Tianjin, 300071, P.R.\ China,}\\
and \textit{Merton College, University of Oxford, OX1 4JD, UK}
\vskip0.25cm

${}^c$\textit{Centre for Theoretical Physics, NITheP, and School of Physics,\\
University of the Witwatersrand, Johannesburg, WITS 2050, South Africa}
\vskip0.25cm

${}^d$\textit{Department of Physics, Northeastern University, Boston, MA 02115, USA}\\
and \textit{ICTP, Strada Costiera 11, Trieste 34014, Italy}

\end{center}

\vskip0.5cm

\begin{abstract}
We explain the origin of the Veronese surface in the vacuum moduli space geometry of the MSSM electroweak sector.
While this result appeared many years ago using techniques of computational algebraic geometry, it has never been demonstrated analytically.
Here, we present an analytical derivation of the vacuum geometry of the electroweak theory by understanding how the F- and D-term relations lead to the Veronese surface.
We moreover give a detailed description of this geometry, realising an extra branch as a zero-dimensional point when quadratic Higgs lifting deformations are incorporated into the superpotential.
\end{abstract}

\newpage

\renewcommand{\thepage}{\arabic{page}}
\setcounter{page}{1}
\def\thefootnote{\arabic{footnote}}
\setcounter{footnote}{0}

\section{Introduction and summary}
The scalar potential of the Minimal Supersymmetric Standard Model (MSSM) is nearly flat along many directions in field space.
The effort to understand the vacuum moduli space geometry, which is the solution to F-flatness and D-flatness constraints on the ${\cal N}=1$ supersymmetric gauge theory, was pioneered in the work of Gherghetta, Kolda, and Martin~\cite{Gherghetta:1995dv}.
Determining this geometry is an important open problem because knowing this structure may facilitate the construction of string and D-brane models for particle physics~\cite{dm}.
Despite many attempts to solve for the vacuum moduli space of the MSSM, a full characterization of the geometry remains elusive.

The complexity of the problem enforced the use of computational techniques in algebraic geometry such as various Gr\"obner basis algorithms and led to several, already striking, partial results.
(See, for example,~\cite{Gray:2009fy} for a physicist's introduction to Gr\"obner bases.)
For instance, the appearance of the Veronese surface in the electroweak sector as an almost generic vacuum moduli space geometry was reported in~\cite{us, Gray:2006jb}.
However, the computing power required for solving Gr\"obner basis problems of this type typically exceeds what is feasible on desktop computers, and it has been difficult to push the analysis past the electroweak sector.
This is because the number of gauge invariant operators in the full theory is more than an order of magnitude larger than in the electroweak sector.

In order to overcome the computational complexity of the problem, Molien integrals and Hilbert series have been used to provide deeper insight into the vacuum moduli space geometry, in particular its dimension.
The plethystic logarithm provides a way to count the number of relations and syzygies among gauge invariant operators (GIOs), and hence it allows the computation of the dimensionality of the vacuum moduli space~\cite{Benvenuti:2006qr}.
Results have been obtained for supersymmetric QCD~\cite{Gray:2008yu} and flavour invariant theories~\cite{Hanany:2010vu}.
Recent investigations have applied the machinery of numerical algebraic geometry to the problem~\cite{Mehta:2012wk, Hauenstein:2012xs}.
Promising advances in this direction allow us to envisage the use of supercomputers to calculate the full MSSM moduli space in a few years' time.

In this work, we shall examine the vacuum moduli space \textit{analytically} without reliance on the computational algebraic geometry packages.
We demonstrate the analytic origin of the relations leading to the Veronese geometry in the electroweak sector of the MSSM, thus completing and justifying the previous results in the literature.
This is an initial step.
We believe that the analytic approach in concert with improved computational techniques will enhance our understanding of the MSSM vacuum geometry.

A general $\cN=1$ globally supersymmetric action in four dimensions is given by the action
\bea
S &=& \int d^4x\ \Bigg[ \int d^4\theta\ \Phi_i^\dagger e^V \Phi_i \;+ 
\Bigg(\frac{1}{4g^2} \int d^2\theta\ \tr{W_\alpha W^\alpha} + \int d^2\theta\ W(\Phi) + {\rm h.c.} \Bigg) \Bigg]\;,
\eea
where $\Phi_i$ are chiral superfields, $V$ is a vector superfield, $W_a$ are chiral spinor superfields, and $W$ is the superpotential given by a holomorphic function of the $\Phi_i$. We imagine there being $n$ such fields $\Phi_i$, such that $i=1, \ldots, n$.
Each of these objects transform under the action of the gauge group $G$ of the theory:
$\Phi_i$ is in some representation $R_i$, and the vector $V$ transforms in the Lie algebra $\cg$ of $G$.
The chiral spinor superfields are the gauge field strength of the theory and are given by $W_\alpha = i\overline{D}^2 e^{-V} D_\alpha e^V$.

The vacuum moduli space is obtained for the expectation values $\phi_{i 0}$ of the scalar component of the superfields $\Phi_i$, satisfying the F-term equations:
\be
\label{fterm}
\left. \frac{\partial W(\phi)}{\partial \phi_i}
\right|_{\phi_i=\phi_{i 0}} = 0 
\ee
and the D-term equations:
\be
\label{dterm}
D^A = \sum_i \phi_{i 0}^\dagger\, T^A\, \phi_{i 0} = 0 ~.
\ee
Here, $T^A$ are generators of the gauge group in the adjoint representation, and we have chosen the Wess--Zumino gauge.
For every solution to the F-flatness conditions, there is one and only one solution to the D-flatness constraints.
Therefore, the latter can be thought of as a gauge fixing condition and the vacuum moduli space corresponds, as an algebraic variety, to the symplectic quotient of all the F-term solutions by the complexified gauge group $G^c$.
For more details, we refer the reader to the literature~\cite{lt, Buccella:1982nx, Gatto:1986bt, Procesi:hr, witten93}; see also,~\cite{Gray:2006jb, Hauenstein:2012xs} for an overview.

Algebraic geometry is a useful and powerful tool to tackle problems in gauge field theories, not least the task of describing the vacuum moduli spaces, which are solutions to polynomial equations in the fields.
Recently, it has been shown that the problem of solving~\eqref{fterm} and~\eqref{dterm} is equivalent to an elimination algorithm~\cite{Hauenstein:2012xs}.
Let us denote the gauge invariant operators by $r_j( \{ \phi_i \})$ with $j=1, \ldots, k$. While the set of all gauge invariant operators is infinite, it will be generated by a minimal set of $k$ generators, and we here take $r_j( \{ \phi_i \})$ to represent that set. Thus, for example, we will find that $k=22$ for the electroweak sector of the MSSM, but $k=991$ for the full MSSM.
We consider the ideal
\begin{equation}
\left\langle
\frac{\partial {W}}{\partial \phi_i} \ ,
y_j - r_j( \{ \phi_i \})
\right\rangle
\subset
R = \IC[\phi_{i=1, \ldots, n}, y_{j = 1, \ldots, k}] \ ,
\end{equation}
where $y_i$ are additional variables.
Then, eliminating all variables $\phi_i$ of this ideal will give an ideal $\cM$ expressed solely in terms of the new variables $y_i$ that corresponds to the vacuum moduli space as an affine variety in the polynomial ring $S=\IC[y_1, \ldots, y_k]$.

From an algebraic geometry point of view, this algorithm corresponds to finding the image of a ring map $D$ from the quotient ring $\cF=\IC[\phi_1, \ldots, \phi_n] \left/ \left\langle\frac{\partial {W}}{\partial \phi_i}\right\rangle\right.$ to the ring $S$,
\be
\cM \simeq {\rm Im}\left(\cF\quad \stackrel{D=\{r_j( \{ \phi_i \})\}}{\xrightarrow{\hspace*{2cm}}} \quad S \right)\;.
\ee
In other words, the algorithm is equivalent to asking what are the relations among GIOs that satisfy the F-flatness conditions.
This is the strategy we adopt in our calculations.

Our aim is to discuss the output of such an algorithm for the MSSM electroweak sector in order to describe the origin of the corresponding vacuum moduli space geometry.
We will see that the Veronese surface stems from a single class of GIOs, namely the $LLe$ operators.
Their intrinsic relations and syzygies define a five-dimensional toric variety.
Three additional relations imposed by the F-term equations reduce the space further to a Veronese surface.
This happens for a superpotential including right-handed neutrinos and/or quadratic Higgs terms.
We also demonstrate that for the latter case, the quadratic nature of the superpotential terms leads to an additional branch in the vacuum moduli space as an extra point that appears from the solution of the above algorithm, giving a vacuum expectation value (VEV) to the $H\barH$ operator.

The organisation of this letter is as follows.
In Section~2, we present the vacuum geometry of the electroweak sector with only the minimal renormalisable superpotential, obtaining a five-dimensional toric variety.
In Section~3, we consider the addition of right-handed neutrino fields and quadratic Higgs terms in the superpotential separately.
We demonstrate how these additional terms modify the vacuum geometry leading to the Veronese surface.
Section~4 offers some concluding remarks.


\section{Electroweak moduli space}\label{EW-VMS}

The full Minimal Supersymmetric Standard Model (MSSM) is an $\cN=1$ globally supersymmetric gauge theory with gauge group $SU(3)\times SU(2) \times U(1)$.
The symmetry of the electroweak sector corresponds to the $SU(2)\times U(1)$ subgroup of the full theory.
In this work, we will consider the electroweak sector only and will subsequently adopt the notation in Table~\ref{tbl:EW} for the indices and the field content of the theory.
This is equivalent to setting the VEVs of the scalar quark fields in the MSSM to zero.

\begin{table}[thb]
{\begin{center} 
\begin{tabular}{|c|c|}\hline
FIELDS & \\ \hline \hline
$L_{\alpha}^i$  & $SU(2)_L$ doublet leptons \\
$e^i$ & $SU(2)_L$ singlet leptons \\
$H_\alpha$ & up-type Higgs \\
$\barH_\alpha$ & down-type Higgs \\ \hline
\end{tabular}

\vspace{0.2cm}

\begin{tabular}{|c|c|}\hline
INDICES & \\ \hline \hline
$i,j,k,l = 1,2,3$ & Flavour (family) indices \\
$\alpha, \beta, \gamma, \delta = 1,2 $ & $SU(2)_L$ indices \\
\hline
\end{tabular}
\end{center}}
{\caption{\label{tbl:EW} {\bf Index conventions and field content
of the electroweak theory}.}}

\end{table}

We consider the usual three generation model.
The theory consists of $13$ superfields.
For clarity, we will drop indices when the context does not allow any confusion.
As noted in Section~1, the scalar component of each superfield is governed by the scalar potential of the theory, which is flat along many directions in field space.
This work aspires to describe the geometry of the flat directions.
Indeed, this is what we mean when we refer to the vacuum moduli space of the MSSM electroweak sector.

As our starting point, let us consider the minimal renormalisable superpotential consistent with R-parity conservation,
\bea
W_{\rm minimal} &=& \; C^0 \sum_{\alpha, \beta} H_\alpha \barH_\beta \eps  + \label{renorm} \sum_{i,j}C^3_{ij} e^i \sum_{\alpha, \beta} L^j_{\alpha} \barH_\beta \eps ~,
\eea
where we have designated coupling constants by $C$ and $\eps$ is the totally antisymmetric tensor. The superpotential~(\ref{renorm}) is precisely the electroweak sector of the MSSM, in the absence of right-handed neutrino superfields. R-parity is defined as $R=(-1)^{3(B-L)+2s}$, where $B$ and $L$ are the baryon and lepton numbers of the superfield, and $s$ is the spin of each component field. It is postulated as a conserved quantum number of the full MSSM superpotential in order to ensure the stability of the proton. The problem of finding the vacuum moduli space of this theory is equivalent to solving the F-term equations~\eqref{fterm} and D-term equations~\eqref{dterm} with this superpotential.

First, let us write down the F-terms explicitly.
These are given by:
\bea\label{expl-F-min1}
&&\frac{\partial W_{\rm minimal}}{\partial H_\alpha}= C^0 \barH_\beta \eps ~,\\
&&\frac{\partial W_{\rm minimal}}{\partial \barH_\beta}= C^0 H_\alpha \eps + \sum_{i,j} C^3_{ij} e^i L^j_{\alpha} \eps ~,\\
&&\frac{\partial W_{\rm minimal}}{\partial L^j_{\alpha}}= C^3_{ij} e^i \barH_\beta \eps ~, \\\label{expl-F-min4}
&&\frac{\partial W_{\rm minimal}}{\partial e^i}= C^3_{ij} \sum_{\alpha, \beta} L^j_{\alpha} \barH_\beta \eps ~.
\eea
The flatness condition requires these terms to vanish and therefore implies the following constraints from the F$_{H_\alpha}$- and F$_{\barH_\beta}$-terms, respectively:
\bea\label{minflatHb}
\barH_\beta&=&0\;,\\\label{minflatH}
C^0 H_\alpha + \sum_{i,j}C^3_{ij}e^iL^j_{\alpha}&=&0\;.
\eea
The other two F-term equations (for the $L$ and $e$ fields) do not lead to any extra constraints as the vanishing of the $\barH$ fields render them trivial.

We now need to tackle the D-flatness conditions \eqref{dterm}, and we will adopt the strategy presented in the introduction.
We look for the space of all holomorphic GIOs built out of F-flat field configurations.
The vacuum moduli space will correspond to an affine variety in $\IC^{22}$ given by an ideal of $S=\IC[y_1, \ldots, y_{22}]$ with $\{y_j=r_j(\{\Phi_i\})\}$, where $r_j$ is a minimal generating set of GIOs.
The $22$ counts the number of GIOs for the electroweak theory under consideration.
A minimal complete set of operators is listed in Table~\ref{gio-ew}.
\begin{table}[h]
{\begin{center}
\begin{tabular}{|c||c|c|c|}\hline
\mbox{Type} & \mbox{Explicit Sum} & \mbox{Index} & \mbox{Number} \\
\hline \hline
$LH$  & $L^i_\alpha H_\beta \eps$ & $i=1,2,3$ & $3$ \\ \hline
$H\barH$ & $H_\alpha \barH_\beta \eps$ & & 1  \\ \hline
$LLe$ & $L^i_\alpha L^j_\beta e^k \eps$ & $i,k=1,2,3;
j=1,2$ & $9$  \\ \hline
$L\barH e$ & $L^i_\alpha \barH_\beta \eps e^j$ & $i,j=1,2,3$ & $9$ \\
\hline
\end{tabular}
\end{center}}{\caption{\label{gio-ew}{\bf Generators of the GIOs for the electroweak sector of the MSSM}.}}
\end{table}

We have already established that the $\barH$ fields vanish by virtue of \eqref{minflatHb}.
The operators containing such fields must consequently vanish in the vacuum, and we are left with $LH$ and $LLe$.
The second constraints \eqref{minflatH} give us relations between these two types of operators.
Indeed, contracting this equation by $L^i_\beta \eps$, we obtain
\be
C^0 L^i_\alpha H_\beta \eps+\sum_{j,k}C^3_{jk}L^i_\alpha L^j_{\beta}e^k\eps=0 \;.
\ee
There will be one such linear equation per $LH$, and the corresponding values for these operators in the vacuum will be completely determined the $LLe$ operators.
Therefore the vacuum moduli space degrees of freedom are the $LLe$ variables only, and with an adequate labeling choice of the $y$ coordinates, the moduli space geometry reduces to an affine variety in $\IC[y_1, \ldots, y_9]$ given by the relations among the $LLe$ polynomials.
The remaining three $y$ coordinates resulting from the $LH$ operators simply provide a linear embedding onto the bigger ring $\IC[y_1, \ldots, y_{12}]$.

Consequently, understanding the relations among the $LLe$ operators is crucial to characterising the geometry of the vacuum moduli space.
These operators are products of the three fields $e^i$ and the three terms $L^i_\alpha L^j_\beta \eps$.
Clearly, they will be subject to the relations,
\be\label{LLe-rel}
(L^i_\alpha L^j_\beta e^k \eps)(L^m_\alpha L^n_\beta e^p \eps)=(L^m_\alpha L^n_\beta e^k \eps)(L^i_\alpha L^j_\beta e^p \eps) \;,
\ee
as can easily be seen by division.
In other words, a set of operators with a common $e^i$ field will be linearly proportional to another set of operators with a common $e^j$ field ($i \neq j$).
Let us introduce the following convenient labeling:
\be\label{y-label}
y_{i+j-2+3(k-1)}=L^i_\alpha L^j_\beta e^k \eps\;.
\ee
With this notation, the relations (\ref{LLe-rel}) can be written as an ideal given by nine quadratic polynomials in the following way:
\bea\label{ideal}\nn
\langle& y_1y_5-y_2y_4,\; y_1y_6-y_3y_4,\; y_2y_6-y_3y_5, & \\ & y_1y_8-y_2y_7,\; y_1y_9-y_3y_7,\; y_2y_9-y_3y_8, & \\ & y_4y_8-y_5y_7,\; y_4y_9-y_6y_7,\; y_5y_9-y_6y_8 & \rangle \;. \nn
\eea
This is a five-dimensional algebraic variety as stated in~\cite{Gray:2006jb}.
(It should be noted that this reference does not give the ideal explicitly.)
Furthermore,~\eqref{ideal} is in fact an irreducible non-compact affine toric variety in $\IC^9$ which is Calabi--Yau.
(These assertions result from computations with algebraic geometry packages such as \texttt{ Macaulay~2}~\cite{mac} and \texttt{Singular}~\cite{sing}, and a detailed exposition of these statements is a work in progress~\cite{future}.)


\section{The Veronese surface}

Let us now turn to the study of the origin of the Veronese surface in the vacuum moduli space of the MSSM electroweak sector.
This geometry appears when additional renormalisable terms are included in the superpotential~\eqref{renorm}, resulting in giving masses to certain fields and lifting parts of the vacuum flat directions.
The original results were obtained using techniques in computational algebraic geometry and were reported in~\cite{us, Gray:2006jb}.
Here, we adopt an analytical approach, showing explicitly the origin of the Veronese surface.

We will consider two cases: firstly, the addition of right-handed neutrino fields and then secondly, quadratic Higgs-lifting terms.
The latter has the property of introducing an additional zero-dimensional point in the vacuum moduli space that has not been noticed in previous works.

\subsection{Right-handed neutrinos}

Let us first study the superpotential with the addition of right-handed neutrino fields.
We keep the same conventions and fields as in Table \ref{tbl:EW} and simply add right-handed neutrino fields and the corresponding new generators of GIOs.
These are presented in Table \ref{tbl:Nu}.
\begin{table}[thb]
{\begin{center}
\begin{tabular}{|c|c|}\hline
FIELDS & \\ \hline \hline
$\nu^i$ & $SU(2)_L$ singlet neutrinos \\ \hline
\end{tabular}

\vspace{0.2cm}

\begin{tabular}{|c||c|c|c|}\hline
\mbox{Type} & \mbox{Explicit Sum} & \mbox{Index} & \mbox{Number} \\
\hline \hline
$\nu$  & $\nu^i$ & $i=1,2,3$ & $3$ \\ \hline
\end{tabular}
\end{center}}
{\caption{\label{tbl:Nu} {\bf Right-handed neutrino fields and corresponding generator of GIOs}.}}
\end{table}

The superpotential is modified as follows.
We include the renormalisable terms corresponding to Majorana and Dirac masses:
\be\label{neut-renorm}
W_{\rm neutrinos} = \sum_{i,j}C^4_{ij} \nu^i\nu^j+\sum_{i,j} C^5_{ij} \nu^i\sum_{\alpha, \beta} L^j_{\alpha} H_\beta \eps\;.
\ee
Here, $C^4$ and $C^5$ are new coupling constants.
(While the neutrino fields $\nu^i$ are themselves gauge invariant, we do not include tadpole-like R-parity violating operators into the superpotential.)
The full superpotential will thus be given by the sum of~\eqref{renorm} and~\eqref{neut-renorm}.
From this, we obtain the following F-term equations:
\bea \label{vernH}
\sum_{i,j} C^5_{ij} \nu^i L^j_{\alpha}\eps-C^0 \barH_\alpha \eps&=& 0 ~, \\ \label{vernHb}
C^0 H_\alpha \eps + \sum_{i,j} C^3_{ij} e^i L^j_{\alpha} \eps&=& 0 ~, \\ \label{vernL}
C^5_{ij} \nu^i  H_\beta \eps+C^3_{ij} e^i \barH_\beta \eps&=& 0 ~, \\ \label{vernnu}
C^4_{ij} \nu^j+ C^5_{ij} \sum_{\alpha, \beta} L^j_{\alpha} H_\beta \eps&=& 0 ~, \\ \label{verne}
C^3_{ij} \sum_{\alpha, \beta} L^j_{\alpha} \barH_\beta \eps&=& 0 ~.
\eea
These equations naturally reduce to equations \eqref{expl-F-min1}--\eqref{expl-F-min4} when the coupling constants $C^4$ and $C^5$ are set to zero.
We follow a similar strategy as in the previous section to analyse the meaning of these constraints.

A couple of immediate conclusions can be drawn.
Firstly, from (\ref{verne}) and from the non-singularity of the coupling matrix $C^3_{ij}$, we conclude that the $L\barH$ must all vanish.
Secondly, we can contract~(\ref{vernL}) with $L^k_\beta$ to obtain:
\be
C^5_{ij} \nu^i\sum_{\alpha, \beta} L^k_{\alpha} H_\beta \eps+ C^3_{ij} e^i \sum_{\alpha, \beta} L^k_{\alpha} \barH_\beta \eps=0\;.
\ee
The second term vanishes by virtue of $L\barH=0$, and we deduce that
\be
\nu^i\sum_{\alpha, \beta} L^k_{\alpha} H_\beta \eps=0\;.
\ee
This implies that both the $\nu$ and $LH$ operators vanish.
This is easy to show.
If $\nu^i\neq0$ then $\sum_{\alpha, \beta} L^k_{\alpha} H_\beta \eps=0$, and from equation~(\ref{vernnu}) we conclude that $\nu^i=0$, in contradiction of the starting hypothesis. 
Therefore $\nu=0$, which implies $LH=0$ from~(\ref{vernnu}).
Finally, from~(\ref{vernH}), we also have $\barH=0$.
To sum up, the following types of GIOs vanish in the vacuum:
\bea
\nu&=&0 ~, \\
LH&=&0 ~, \\
H\barH&=&0 ~, \\
L\barH e&=&0 ~.
\eea

The only non-trivial operators are therefore the $LLe$.
Moreover, the only non-trivial F-term equation remaining is \eqref{vernHb}, which is similar to the result of previous section.
The major difference is that now, the $LH$ operators must vanish.
Consequently, contracting~\eqref{vernHb} with $L^k_\beta$, we obtain:
\be\label{con}
\sum_{i,j} C^3_{ij} e^i L^j_{\alpha}L^k_\beta \eps= 0 \;.
\ee
This condition is the extra condition implied by the new right-handed neutrino terms in the superpotential.
Naturally, we still retain the intrinsic relations among $LLe$ operators that plays a crucial role in determining the vacuum moduli space geometry.
We will see in the next subsection that equation~\eqref{con} forces the vacuum geometry to be a Veronese surface.

\subsection{Veronese geometry and $LLe$ operators}

The relations of the $LLe$ operators still hold and the geometry is given by the constraints~(\ref{LLe-rel}) together with the above extra conditions~\eqref{con}.
It turns out that these are precisely the definition of a Veronese surface.
To see this, let us introduce new electron variables in the following way:
\be
 \eB^j\equiv \sum_iC^3_{ij} e^i.
\ee
These new variables are as good variables as $e^i$ if the matrix $C^3_{ij}$ is non-singular.
Let us write the corresponding GIOs variables as $\yB$ according to the following conventions:
\be
\yB_{i+j-2+3(k-1)}=(-1)^{k-1}\sum_{\alpha,\beta}L^i_\alpha L^j_\beta \eB^k \eps\;,\quad {\rm for} \; i<j\;.
\ee
With these new definitions, the ideal~(\ref{ideal}) maintains the same structure and the constraints (\ref{con}) are equivalent to the following:
\bea\label{Veronese}
\yB_1-\yB_9=0 ~, \\
\yB_2-\yB_6=0 ~, \\
\yB_4-\yB_8=0 ~.
\eea
Therefore the full ideal is given by:
\bea\label{new-ideal}\nn
\langle \quad \yB_1\yB_5-\yB_2\yB_4,\; \yB_1\yB_6-\yB_3\yB_4,\; \yB_2\yB_6-\yB_3\yB_5,\\\nn \yB_1\yB_8-\yB_2\yB_7,\; \yB_1\yB_9-\yB_3\yB_7,\; \yB_2\yB_9-\yB_3\yB_8,\\\nn \yB_4\yB_8-\yB_5\yB_7,\; \yB_4\yB_9-\yB_6\yB_7,\; \yB_5\yB_9-\yB_6\yB_8,\\
\yB_1-\yB_9,\; \yB_2-\yB_6,\; \yB_4-\yB_8 \quad \rangle \;.
\eea
This ideal defines a three-dimensional algebraic variety in $\IC^9$.

To see why this set of conditions corresponds to the Veronese surface, let us recall the definition.
The Veronese surface is an embedding of $\IP^2$ into $\IP^5$ given by:
\be
\ba{ccc}
\IP^{2} & \rightarrow & \IP^{5} \cr
[x_0:x_1:x_2] & \mapsto &
[{x_0}^2 : x_0x_1 : {x_1}^2 : x_0x_2 : x_1x_2 : {x_2}^2]\label{veronese}
\ea
\ee
Let us introduce the following change of $LL\eB$ operators variables for the three degrees of freedom of the variety defined by \eqref{new-ideal}:
\bea
\yB_3&\rightarrow x_0^2 ~, \\
\yB_5&\rightarrow x_1^2 ~, \\
\yB_7&\rightarrow x_2^2 ~.
\eea
This is a covering map changing the multiplicity of the variety but not the geometry.
From this, the change of variables for all the remaining $\yB$ is defined.
Indeed, we can consider the following sub-ideal
\bea
\langle & \yB_2\yB_6-\yB_3\yB_5,\; \yB_2-\yB_6 & \rangle\;,
\eea
which leads to the constraints $\yB_2^2=\yB_6^2=\yB_3\yB_5$, and therefore $\yB_2=\yB_6=x_0x_1$.
Similarly, for all the remaining variables, we conclude that the polynomial relations from the ideal \eqref{new-ideal} leads to the full change of variables:
\be\label{verovar}
\ba{lll}
\yB_1 \rightarrow x_0x_2 ~, & \yB_2 \rightarrow x_0x_1 ~, & \yB_3 \rightarrow x_0^2 ~, \\
\yB_4 \rightarrow x_1x_2 ~, & \yB_5 \rightarrow x_1^2 ~, & \yB_6 \rightarrow x_1x_0 ~, \\
\yB_7 \rightarrow x_2^2 ~, & \yB_8 \rightarrow x_2x_1 ~, & \yB_9 \rightarrow x_2x_0 ~.
\ea
\ee
This change of variables gives the mapping between the various defining equations for the Veronese surface, the ideal~\eqref{new-ideal} and the mapping~\eqref{veronese}.
When we projectivise the space $[\yB_3:\yB_5:\yB_7]$, we effectively obtain the corresponding mapping between $\IP^2$ and $\IP^5$.
Thus, we have derived the Veronese surface as the vacuum moduli space analytically.

\subsection{Quadratic Higgs-lifting terms}

The Veronese surface did not appear in~\cite{us, Gray:2006jb} solely for the case of right-handed neutrinos fields.
To illustrate another example, let us consider a case that is radically different, without the introduction of additional fields.
Instead, we will add quadratic terms for the Higgs field in the minimal superpotential~\eqref{renorm}.
We consider the following extra terms:
\be\label{Higgs-sup}
W_{\rm quadratic} = \; \lambda (H_\alpha \barH_\beta \eps)^2 + \lambda_{ij}( L^i_{\alpha} H_\beta \eps)( L^j_{\gamma} H_{\delta} \epsilon^{\gamma\delta})\;,
\ee
where $\lambda$  and $\lambda_{ij}$ are coupling constants.
The two terms in~(\ref{Higgs-sup}) are the only terms allowed by gauge invariance and R-parity conservation at this mass order in the superpotential.
The full superpotential will thus be given by the sum of~\eqref{renorm} and~\eqref{Higgs-sup}.
The corresponding F-terms equations are:
\bea
2\lambda^2 H_{\gamma} \barH_{\delta} \epsilon^{\gamma\delta} \barH_\beta\eps + \label{verHH} C^0 \barH_\beta \eps- 2\lambda_{ij} L^i_{\beta}\eps ( L^j_{\gamma} H_{\delta} \epsilon^{\gamma\delta})&=&0 ~, \\
2 \lambda^2 H_{\gamma} \barH_{\delta} \epsilon^{\gamma\delta} H_\alpha\eps + \label{verHHb} C^0 H_\alpha \eps + \sum_{i,j} C^3_{ij} e^i L^j_{\alpha} \eps&=&0 ~, \\\label{verHL}
2 \lambda_{ij} H_\beta \eps( L^j_{\gamma} H_{\delta} \epsilon^{\gamma\delta})+ C^3_{ij} e^i \barH_\beta \eps&=&0 ~, \\\label{verHe} 
C^3_{ij} \sum_{\alpha, \beta} L^j_{\alpha} \barH_\beta \eps&=&0 ~.
\eea
Again, with vanishing $\lambda$ and $\lambda_{ij}$, these reduce to equations~\eqref{expl-F-min1}--\eqref{expl-F-min4}.

From~\eqref{verHe}, we see that the $L\barH$ operators vanish.
As in the neutrino case, we can contract \eqref{verHL} with $L^k_\alpha$ to obtain:
\be
2 \lambda_{ij} (L^k_\alpha H_\beta \eps)( L^j_{\alpha} H_\beta \eps)+C^3_{ij} e^i L^k_\alpha\barH_\beta \eps=0\;.
\ee
With  vanishing $L\barH$, the second term disappears which implies that $LH$ vanish as well as the constants $\lambda_{ij}$ are non-singular.
We may now contract (\ref{verHHb}) with $L^k_\beta$ to obtain the same condition as in the neutrinos case~\eqref{con}:
\be\label{conH}
\sum_{i,j} C^3_{ij} e^i L^j_{\alpha}L^k_\beta \eps= 0\;.
\ee
We thus have exactly the same conditions for the $LLe$ operators as in the previous subsection in which we considered right-handed neutrinos.

We must still investigate whether the $H\barH$ operators from equation~\eqref{verHH} modify this geometry.
Contracting~\eqref{verHH} with $\barH_\beta$ (alternatively contracting (\ref{verHHb}) with $H_\alpha$) and keeping in mind that $L\barH=0$ (respectively $LH=0$), we obtain the condition:
\be
H_\alpha \barH_\beta \eps(H_\alpha \barH_\beta \eps+C^0/2\lambda^2)=0\;.
\ee
The corresponding solutions are obtained as follows:
\be
H_\alpha \barH_\beta \eps=0\;,\quad {\rm or}\quad H_\alpha \barH_\beta \eps=-C^0/2\lambda^2 \;.
\ee
We thus have two cases.
For the first solution $H\barH$ vanishes, and we only have the Veronese surface from the $LLe$ operators with all remaining GIOs vanishing, exactly in the same way as for the case with right-handed neutrinos.

However, for the other solution with $H_\alpha \barH_\beta \eps=-C^0/2\lambda^2$, clearly $\barH$ cannot equal zero.
However, from~(\ref{verHL}) and vanishing $LH$, we obtain
\be
e^i\barH_\alpha=0\;.
\ee
Consequently, when $\barH\neq0$, we must have $e^i=0$, and therefore, $LLe=0$.
This solution consists then of the point
\be\label{point}
LH=LLe=L\barH e=0 \quad {\rm and} \quad H\barH=-C^0/2\lambda^2\;.
\ee
The full moduli space is constituted of two branches, the Veronese surface as presented in the previous subsection and the single point \eqref{point}. The reader will immediately recognize the latter as precisely the Higgs minimum which spontaneously breaks electroweak symmetry in the Standard Model, with the correct dependence on the quadratic and quartic coefficients in the Higgs potential.

When both Higgs lifting terms~\eqref{Higgs-sup} and right handed neutrinos terms~\eqref{neut-renorm} are taken into consideration for the full superpotential, it is fairly straightforward to realise that the result remains the same, with the moduli space comprised of the above two branches. This is also not surprising, as the second term in~(\ref{Higgs-sup}) can be obtained from those in~(\ref{neut-renorm}) upon integrating out the right-handed neutrino superfields.
It is significant that the Veronese geometry {\em requires} the existence of a Majorana mass term for the right-handed neutrinos. If neutrinos are purely Dirac fermions, with $C_{ij}^4=0$ in~(\ref{neut-renorm}), then the Veronese geometry is not obtained.

This is the main analytic result of this investigation.
As promised, the Veronese surface is the vacuum moduli space of the electroweak sector of the MSSM given phenomenologically realistic superpotentials at renormalisable mass level.


\section{Discussion and outlook}\label{conclusion}

We have unveiled the crucial role of the $LLe$ GIOs in the geometry of the electroweak moduli space.
Their relations and syzygies define a five-dimensional toric variety for the case of a minimal renormalisable superpotential.
When extra terms are included, such as right-handed neutrino operators or quadratic Higgs terms, extra constraints on this variety lead to the Veronese surface.
This solution emerges in a similar way for two very different theories, and it seems reasonable to anticipate that this is the way the Veronese geometry appears in the vacuum moduli space of every case found in~\cite{Gray:2006jb}.

In addition, an extra point in the moduli space has been found for the case of quadratic Higgs terms.
This leads to a disconnected vacuum geometry, and we can expect non-trivial topological solutions from this theory, such as domain walls.
This extra point previously escaped attention from computations due to its zero-dimensionality.

A more complete investigation resulting from a systematic scanning of $\cN=1$ supersymmetric gauge theories, including different number of generations and various superpotential terms is currently being undertaken~\cite{future}.
The promising analytical approach combined with the power of computational algebraic geometry packages lead to reasonable hopes that the vacuum moduli space for different sectors of the MSSM will eventually be written down.
Indeed, a complete description of the MSSM vacuum moduli space might be obtained from a combinations of techniques and with the help of numerical algebraic geometry and supercomputers.
This is for the future.


\section*{Acknowledgements}
We thank James Gray, Jonathan Hauenstein, and Dhagash Mehta for collaborations on similar themes.
VJ is supported by the South African Research Chairs Initiative of the Department of Science and Technology and National Research Foundation.
The work of YHH, VJ, and BDN is partially funded by the US National Science Foundation under the grant CCF-1048082, EAGER: CiC: A String Cartography.

\end{document}